\documentclass[aps,pra,10pt,twocolumn,superscriptaddress,floatfix]{revtex4-1}

\usepackage[T1]{fontenc}
\usepackage[utf8]{inputenc}
\usepackage[english]{babel}

\usepackage{amsmath,amssymb}
\usepackage{bm}

\usepackage{graphicx}

\usepackage{siunitx}
\usepackage{xcolor}
\usepackage[normalem]{ulem}

\usepackage[
  hidelinks,
  colorlinks = true,
  linkcolor  = blue,
  citecolor  = blue,
  urlcolor   = blue
]{hyperref}
\usepackage{cleveref}

\newcommand{\Nt}{N_\mathrm{tr}}
\newcommand{\ntr}{n_\mathrm{tr}}

\newcommand{\tnr}{\tau_\mathrm{nr}}
\newcommand{\trad}{\tau_\mathrm{rad}}
\newcommand{\tp}{\tau_\mathrm{ph}}

\newcommand{\nsat}{n_\mathrm{sat}}
\newcommand{\pth}{p_\mathrm{th}}
\newcommand{\Pcw}{P_\mathrm{cw}}

\begin{document}

%\title{Phonon modulation via multimode self-sustained optomechanical dynamics}
\title{Time-Resolved dynamics of semiconductor nanolaser via four-wave mixing gating}

\author{Federico Monti}
\affiliation{Université Paris-Saclay, CNRS, Centre de Nanosciences et de Nanotechnologies, 91120 Palaiseau, France}

\author{Guilhem Madiot}
\affiliation{Institut de Physique de Nice, CNRS, Université Côte d'Azur, 17 rue Julien Lauprêtre, 06000 Nice, France}

\author{Giuseppe Modica}
\affiliation{Université Paris-Saclay, CNRS, Centre de Nanosciences et de Nanotechnologies, 91120 Palaiseau, France}

\author{Grégoire Beaudoin}
\affiliation{Université Paris-Saclay, CNRS, Centre de Nanosciences et de Nanotechnologies, 91120 Palaiseau, France}

\author{Konstantinos Pantzas}
\affiliation{Université Paris-Saclay, CNRS, Centre de Nanosciences et de Nanotechnologies, 91120 Palaiseau, France}

\author{Isabelle Sagnes}
\affiliation{Université Paris-Saclay, CNRS, Centre de Nanosciences et de Nanotechnologies, 91120 Palaiseau, France}

\author{Alejandro M. Yacomotti}
\affiliation{Université Paris-Saclay, CNRS, Centre de Nanosciences et de Nanotechnologies, 91120 Palaiseau, France}
\affiliation{Laboratoire Photonique Numérique et Nanosciences, Institut d’Optique d’Aquitaine, Université Bordeaux, CNRS, 33405 Talence, France}

\author{Fabrice Raineri}
\affiliation{Université Paris-Saclay, CNRS, Centre de Nanosciences et de Nanotechnologies, 91120 Palaiseau, France}
\affiliation{Institut de Physique de Nice, CNRS, Université Côte d'Azur, 17 rue Julien Lauprêtre, 06000 Nice, France}
\affiliation{fabrice.raineri@univ-cotedazur.fr}

\begin{abstract}
We experimentally demonstrate the direct time-domain characterization of photonic-crystal nanolasers at telecom wavelengths using a nonlinear optical gating technique based on four-wave mixing. This approach enables the temporal characterization of the ultrafast emission dynamics under short-pulse excitation with picosecond time resolution. When a weak continuous-wave component is added to the pulsed pump, the emission becomes deterministic and the build-up time is considerably reduced. The difference between purely pulsed and hybrid excitation regimes points to the influence of pulse-to-pulse timing fluctuations. To elucidate this effect, we perform Langevin-based simulations that reproduce the experimentally observed broadening and confirm that time jitter, originating from spontaneous-emission noise near threshold, dominates the temporal dispersion. These results establish four-wave-mixing gating as a powerful method to probe nanolaser dynamics with picosecond precision.
\end{abstract}
\maketitle

\section{Introduction}

Future photonic devices are expected to operate with energy consumption on the order of 1 fJ/bit, achieving bandwidths exceeding 10 GHz and footprints below \qty{100}{\micro\meter^2} \cite{Miller2009}. Among the various candidate light sources, photonic crystal (PhC) nanolasers are particularly promising, as they combine diffraction-limited modal volumes $(\lambda/2n)^3$ \cite{Zhang2010} with high quality factors ($10^5$–$10^7$) \cite{Crosnier2016}. Such strong confinement enhances light–matter interaction, thereby accelerating spontaneous emission in the lasing mode and consequently reducing the threshold. For this reason, PhC nanolasers are expected to exhibit ultrafast emission on the tens-of-picoseconds timescale \cite{Yacomotti2013,Miller2009,bazin2014ultrafast}, with typical peak powers of about \qty{1}{\micro\watt} \cite{Bjork1991}.
Beyond applications, nanolasers are also of fundamental interest. In the high-$\beta$ regime, the conventional nonlinear input–output curve softens and, as $\beta \rightarrow 1$, becomes nearly linear \cite{Strauf2006,Deng2021}, challenging the traditional notion of a laser threshold and giving rise to so-called thresholdless lasers \cite{khajavikhan_thresholdless_2012,prieto_near_2015}. This regime also makes the onset of lasing increasingly noise-dominated. As a result, accessing the build-up dynamics and timing fluctuations becomes crucial for properly identifying the threshold and understanding the role of spontaneous emission.

Despite their potential, detailed investigations of nanolaser emission dynamics at telecom wavelengths (1550 nm) remain limited by detector performance, both in terms of sensitivity and speed. In the visible and ultraviolet ranges, silicon avalanche photodiodes (Si-APDs) achieve efficiencies near 70\% and dark count rates as low as $\sim$100 s$^{-1}$ \cite{Salzer2008}, but their sensitivity drops sharply in the infrared. Furthermore, most IR detectors suffer from high noise unless operated at cryogenic temperatures \cite{Ma2009}. Streak cameras, while providing sub-picosecond resolution and single-photon sensitivity in the visible range, exhibit a significant efficiency drop in the near-infrared.
For these reasons, most studies on nanolasers have focused on their steady-state properties, typically characterized under continuous-wave (CW) operation. 

Yet, accessing the ultrafast dynamics of semiconductor nanolasers requires time-resolved techniques capable of resolving the emission build-up following a short excitation. Early studies of vertical-cavity and 2D photonic-crystal lasers demonstrated that the relevant timescales of the associated temporal profiles lie in the tens-of-picoseconds range and cannot be accurately obtained from steady-state measurements alone \cite{shah2013ultrafast}. To overcome detector limitations in the telecommunication band, nonlinear optical gating techniques based on frequency up-conversion were introduced, enabling direct reconstruction of nanolaser temporal responses under femtosecond excitation, down to sub-picosecond resolution  \cite{shah1987subpicosecond, bouche1998dynamics, Raineri2009}. These works established that the laser turn-on time decreases with increasing excitation strength and that the temporal response is strongly asymmetric, reflecting the coupled carrier–photon dynamics triggered by an ultrashort pump pulse. Importantly, such approaches provide a direct measurement of the emission dynamics without relying on electronic bandwidth, making nonlinear optical gating a uniquely suited tool to investigate ultrafast nanolaser operation at telecommunications wavelengths.

\begin{figure}[!ht] 
\centering
\includegraphics{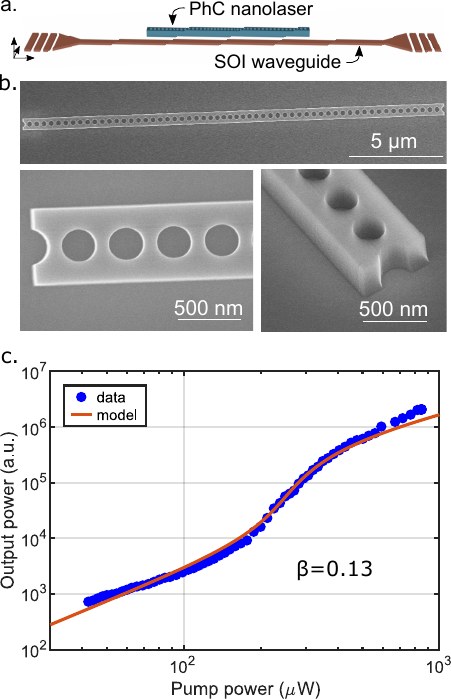} 
\caption{\textbf{a.} Schematics of the integrated nanolaser. \textbf{b.} SEM images of the photonic-crystal nanolaser. Top-view of the full PhC nanobeam (top), zoom-in top-view (left) and angled-view (right) of the nanobeam ending. \textbf{c.} Characterization of the laser curve under CW pumping. Experimental data (blue squares) and simulation obtained from the calibrated model (red), where $\beta=0.13$.}
\label{fig1}
\end{figure}

Here, we present a detection technique based on four-wave mixing (FWM) that enables direct observation of the transient response of 1D nanocavity lasers under a kick-like pumping. This approach provides access to fundamental dynamical quantities such as build-up time and ultimate modulation speed, offering an unprecedentedly complete picture of nanolaser operation in the time domain, with 2 ps temporal resolution. 

The nanobeam cavity consists of a 650~nm wide and 285~nm thick InP-based ridge, incorporating four InGaAsP quantum wells. The nanobeam is patterned with a 1D photonic crystal consisting of air holes of radius $r = 120$ nm. The lattice constant is tapered from 300 nm at the cavity center to 330 nm at the edges, allowing the confinement of an electromagnetic mode around $\lambda = 1550$ nm with a Gaussian-like envelope, thereby minimizing out-of-plane radiation losses \cite{Bazin2014}. The coupling to the external waveguide is achieved by heterogeneous integration of the III-V semiconductor layer on a silicon-on-insulator (SOI) waveguide with dimensions of 500 nm × 220 nm. Diffraction gratings at the waveguide terminations enable the collection of the optical signal into an optical fiber (see \cref{fig1}.a). SEM images of the device are shown in ~\cref{fig1}.b.

First, we proceed to the characterization of the laser emission under continuous-wave (CW) pumping. A CW laser diode at 1180 nm is focused on the nanocavity via a microscope objective, and the emission spectrum is analyzed with a spectrometer coupled to a nitrogen-cooled InGaAs camera. ~\Cref{fig1}.c shows the laser curve, which exhibits a characteristic S-shaped behavior \cite{Kapon1999}. The estimated threshold power is approximately \qty{250}{\micro\watt}. The data are shown with the stationary solutions obtained from the numerical integration of the calibrated model, as detailed in the following.

\section{Optical gating experiment}

\begin{figure*}[!ht] 
\centering
\includegraphics[scale = 0.22]{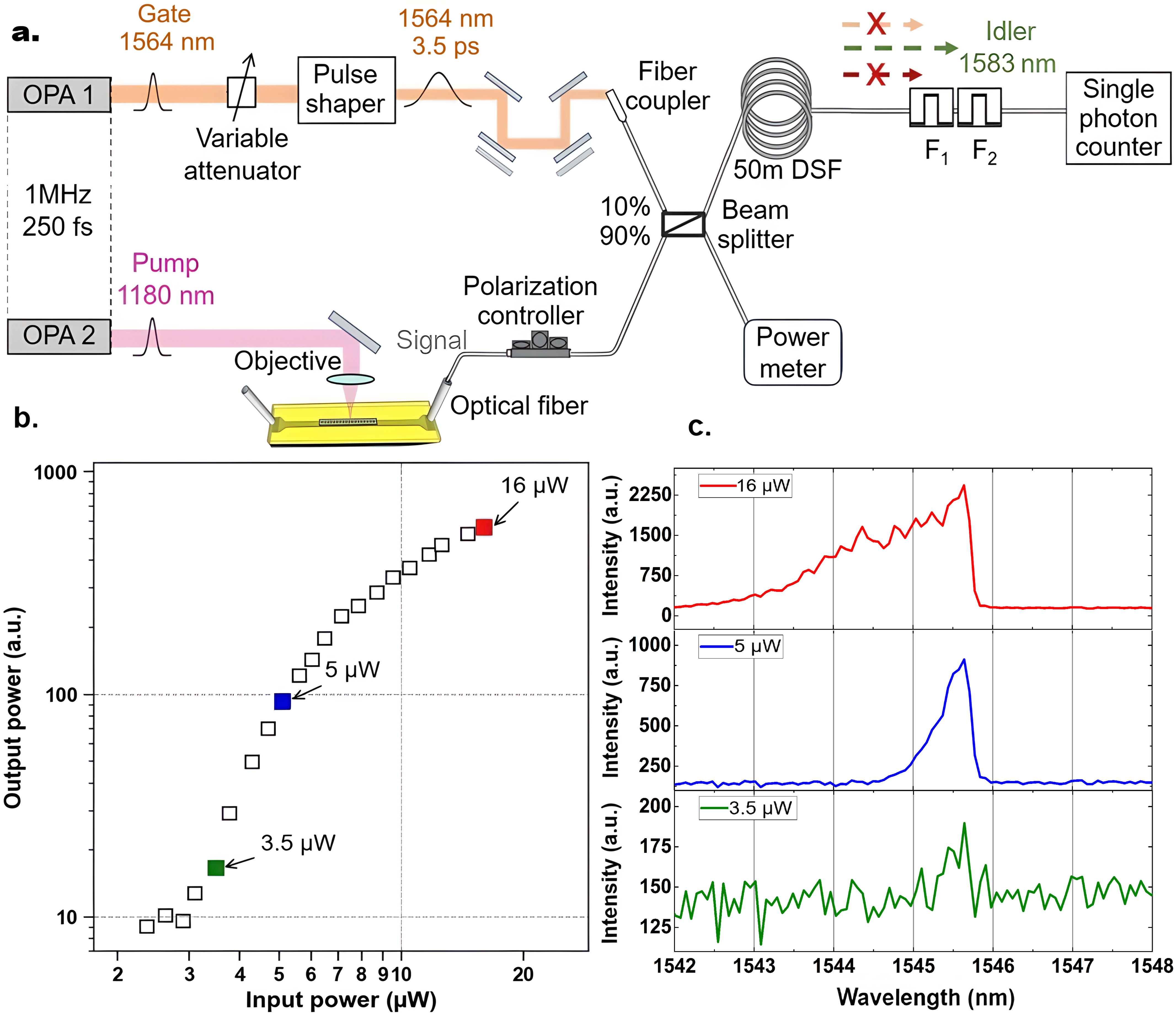} 
\caption{\textbf{Experimental configuration.} \textbf{a.} Schematic of the FWM time-gating experimental setup. \textbf{b.} Measurement of the SOI waveguide output power as a function of the optical power of the pulsed pump, $P_\mathrm{in}$. \textbf{c.} Optical spectra of the output emission for $P_\mathrm{in}=$ \qty{3.5}{\micro\watt} (green), \qty{5}{\micro\watt} (blue), and \qty{16}{\micro\watt} (red). }
\label{fig2}
\end{figure*}

In order to probe the temporal response of the nanolaser, we implement an optical gating experiment based on degenerate pulsed four-wave mixing (FWM). In this scheme, two photons from a pump pulse (hereafter referred to as the gate), at frequency $\omega_g$, interact with one photon from the nanolaser emission (signal), at frequency $\omega_s$, to generate an idler photon at $\omega_i = 2\omega_g - \omega_s$. The nonlinear interaction occurs in a dispersion-shifted fiber (DSF), chosen to minimize group-velocity dispersion near 1550~nm.~\Cref{fig2}.a illustrates the experimental setup used to pump, collect, and detect the nanolaser emission. Two synchronized tunable optical parametric amplifiers (OPA) lasers produce 250~fs pulses at a 1~MHz repetition rate and are split into two branches. 
A total length of 50 meters of DSF is chosen as a tradeoff between input gate power and dispersion to obtain large FWM conversion. A polarization controller is used before the DSF to match the polarization of both the gate and the signal.

The upper branch (OPA~1) provides the gate beam used to sample the nanolaser emission. This beam is spectrally narrowed to $\sim$1~nm using a pulse shaper to mitigate self-phase modulation effects that degrade the signal-to-noise ratio of the FWM process. As a result, the duration of the gate pulse increases to $\sim$3.5~ps. A gate wavelength of 1564~nm is chosen to maximize phase matching into the DSF. Because FWM is a $\chi^{(3)}$ process and since the idler intensity scales with the temporal overlap between the gate and the signal, the effective sampling resolution corresponds to the square root of the gate duration, yielding approximately 1.9~ps~\cite{Agrawal}.
The lower branch (OPA~2) delivers the pumping pulses that generate the transient response from the nanolaser. The emission is collected via an optical fiber and sent through a polarization controller to align the signal polarization with that of the gate beam, thereby maximizing the FWM efficiency. A motorized delay line precisely controls the temporal overlap between the signal and gate pulses. Both beams are then combined using a beam splitter that directs 90\% of the signal and 10\% of the gate into the DSF. The gate peak power in the DSF is approximately 47~W, which is about eleven orders of magnitude higher than the idler.

The idler intensity is measured as a function of the relative delay between the gate and signal pulses, yielding a direct reconstruction of the nanolaser’s temporal profile. To isolate the idler, both the strong gate beam and residual signal photons are suppressed using two identical bandpass filters with a 65~dB extinction ratio, selecting only the idler wavelength at 1583~nm. The filtered photons are then detected by a single-photon counter with a quantum efficiency of 20\%, operated in gated mode in order to reduce the dark counts to a few photons per second.

\section{Results}

Prior to temporal gating, we measure the spectral response of the nanolaser under pulsed excitation. In ~\cref{fig2}.b, we provide three spectra obtained at different pumping pulse powers. For all three spectra, the emission peak is centered near 1545.5~nm. The most distinctive feature is the asymmetric broadening of this peak, which develops a pronounced blue tail that becomes more evident as the pump power increases. This behavior originates from the transient response under pulsed excitation, during which the carrier population varies rapidly over time, inducing wavelength shifts through the free-carrier effect \cite{Soref1987}. In ~\cref{fig2}.c, we show the amplitude of the spectral peak as a function of the pump power, defining a laser curve. Consequently, we estimate the threshold power to  \qty{4.5}{\micro\watt} in the pulsed regime. 

\begin{figure*}[ht] 
\centering
\includegraphics[scale=1]{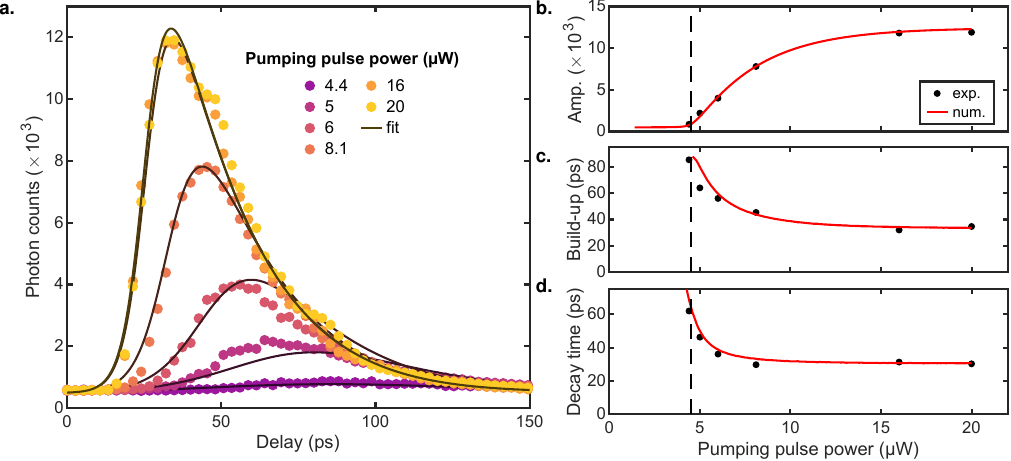} 
\caption{\textbf{Temporal response measurements for increasing pulsed pump powers.} \textbf{a.} The nanolaser is pumped by 250 fs optical pulses.  Data are fitted (gray lines) with the numerical integration of \cref{eq:master}. \textbf{b.} Extracted pulse amplitudes, \textbf{c.} pulses build up time, and \textbf{d.} pulses decay times. The same parameters obtained from the fits are shown in red.}
\label{fig3}
\end{figure*}

In~\cref{fig3}.a, we show the temporal profiles of the nanolaser emission for different pumping pulse powers. The temporal delay is obtained from the delay-line position and is calibrated so that $t = 0$ corresponds to the arrival of the pumping pulse at the sample (see~\cref{ap:delay}). The laser emission does not start immediately after the excitation, but after a delay, known as the build-up time. Then, the profiles exhibit a rapid increase followed by a slower decay, resulting in pronounced asymmetry. 

In \cref{fig3}b-d., we depict the emission peak amplitude (b), build-up time (c), and decay time (d). Above a threshold at approximately \qty{4.5}{\micro\watt}, the peak amplitude grows linearly before saturating, which is consistent with the output power measurement shown in \cref{fig2}.b. The build-up time -- defined as the duration required for the emission to reach its maximum intensity -- decreases from 80 ps to about 34 ps far above the threshold. The pulse decay time is computed from the time trace as the exponential decay time after peaking. It decreases from 60 ps to 30 ps with increasing pumping pulse power.

Under pulsed excitation, higher pump powers generate a larger carrier population, which enhances the recombination rate. 
As a result, the emission starts earlier and peaks sooner, while the carriers also decay more rapidly.
At higher excitation levels -- above \qty{16}{\micro\watt} -- the output power saturates. This behavior arises from the small modal volume of the nanolaser: once the available carrier population is depleted, further increases in pump power no longer create additional carriers, setting an upper limit on the photon population~\cite{Yacomotti2023}.

The experiment is complemented by numerical simulations based on the semiconductor laser rate equations, including a pulsed injection term. The normalized form of these equations, derived as detailed in~\cref{ap:model}, describes the evolution of the photon density $s$ and carrier density $n$ in the nanolaser:
\begin{align} \label{eq:master}
\begin{split}
\dot{s} &= \frac{1}{\tp} \Big( -s + (n-\ntr)s + \beta  n^2 \frac{\tnr }{\trad}  \Big) \\
\dot{n} &= \frac{1}{\tnr} \Big(  R(t)\ (1+\ntr)-n -\frac{n^2 \tnr}{\trad} - (n-\ntr)s \Big)
\end{split}
\end{align} 
where $\tp$ is the photon decay time, $\ntr$ is the carrier density at transparency, $\beta$ is the spontaneous emission factor and $\tnr$ and $\trad$ describe the non-radiative and radiative recombination rates, respectively. The time-dependent injection rate R(t) is modeled as a Gaussian pulse:
\begin{equation}
    R(t) = \Bigg(\frac{\Pcw}{R_0} + \frac{p}{\pth}\frac{ e^{-t^2/\sigma^2}}{\sqrt{\pi \sigma}}\Bigg)\Big(1-\frac{n(t)}{\nsat}\Big)
\end{equation}
where $p$ is the measured pumping pulse power, $\Pcw$ is the CW pump power, $\pth$ and $R_0$ are scaling factors, and $\sigma = \tau_\mathrm{fs}/(2\ln{2})$.
This expression accounts for a saturation effect related to the finite total number of available emitters through the saturation carrier density $\nsat$. 

The experimental data in \cref{fig3}.a is fitted by adjusting the numerical time traces to all the measured temporal responses at once, using $\beta$, $\tp$, $\ntr$, $\nsat$, and $R_0$ as fitting parameters. Other parameters are set to $\tnr=2.6$ ns, as previously reported in \cite{Crosnier2017}, and $\trad=50.7$ ns, as detailed in \cref{ap:model}. The fitting parameters are reported in the left column of \cref{tab:mytable}. 
Using this parameter set, we simulate the temporal profiles as a function of the pumping pulse power and report their peak amplitude, build-time time, and decay time in \cref{fig3}b-c, employing the same methodology as applied to the data. To validate the resulting set of fitting parameters, we simulate a laser curve under CW pumping and compare it with the experimental data in \cref{fig1}.c. 
\begin{figure}[ht] 
\centering
\includegraphics[scale=1]{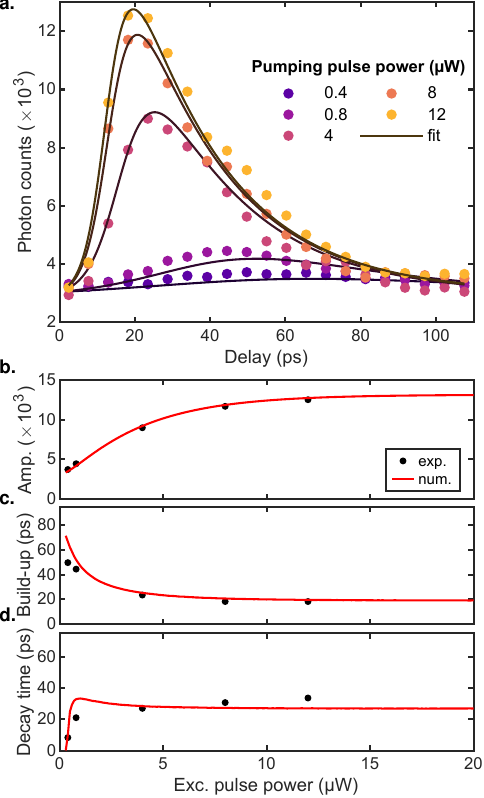} 
\caption{\textbf{Effect of a CW pump power $\Pcw=\qty{250}{\micro\watt}$.} \textbf{a.} The nanolaser is pumped by 250 fs optical pulses.  Data are fitted (gray lines) with the numerical integration of \cref{eq:master}. \textbf{b.} Extracted pulse amplitudes, \textbf{c.} pulses build up time, and \textbf{d.} pulses decay times. The same parameters obtained from the fits are shown in red.}
\label{fig4}
\end{figure}

In this first experiment, a delay of approximately 15 ps is observed before the emission starts. We attribute this time to the exponential build-up of the photon number from the quantum vacuum to saturation. In order to reduce this delay, we reproduce the experiment with a CW pump contribution, setting the CW power to threshold, i.e. with $\qty{250}{\micro\watt}$.
As shown in~\cref{fig4}.a, the resulting temporal profiles peak sooner than those in~\cref{fig3}.a. 
We extract the peak amplitude (a.), build-up time (c.), and decay time (d.) from the data. The peak amplitude increases immediately with the pumping pulse power, i.e. without threshold, as observed in the previous experiment when the CW pump power is turned off. The build-up time decreases quickly from 50 ps to about 20 ps when the pumping pulse power increases. The decay time increases quickly and saturates around 35 ps. 
\begin{table}[h]
\centering
\setlength{\tabcolsep}{15pt} % increase column separation
\begin{tabular}{c c c}
 & without CW & with CW \\
\hline
$\beta$            & $0.13\pm0.03$ & fixed \\
$\ntr$             & $15.2\pm1.1$  & fixed \\
$\nsat$            & $24.0\pm1.5$      & $25.0 \pm 1.7$ \\
$\pth$             & $2.9\pm0.1$   & $2.9\pm0.4$ \\
$\tp$ (ps)              & $26.7\pm0.8$  & $24.3\pm2.6$ \\
$R_0$              &           & $0.026\pm0.007$ \\
$y_\mathrm{scale}$ & $23.5\pm1.5$    & $75.0\pm2.2$ \\
offset             & $5\pm0.3$     & $2.9\pm0.2$ \\
\hline
\end{tabular}
\caption{\textbf{Fit results}. The parameters are obtained by fitting all the experimental time-domain curves at once. The fitting of data taken with or without CW pumping are shown with the associate 95\% confidence interval.}
\label{tab:mytable}
\end{table}

In the first experiment, when $\Pcw=0$, the pulse emission is triggered by spontaneous emission, which must be associated with time jitter, especially near the threshold. Because the photon counter averages over one million pulses, we suspect this effect to artificially broaden the measured temporal profile. Therefore, we fit all the data in \cref{fig4} at once using the model in \cref{eq:master}, using the same methodology as in the previous case, but by fixing $\beta$ and $\ntr$. The resulting set of physical parameters (see \cref{tab:mytable}, right column) is similar to before, except for the photon lifetime, which is smaller but within the confidence interval. %This is expected, as the use of a CW pump reduces time jitter and limits the artificial pulse broadening. 

\begin{figure}[!ht] 
\centering
\includegraphics{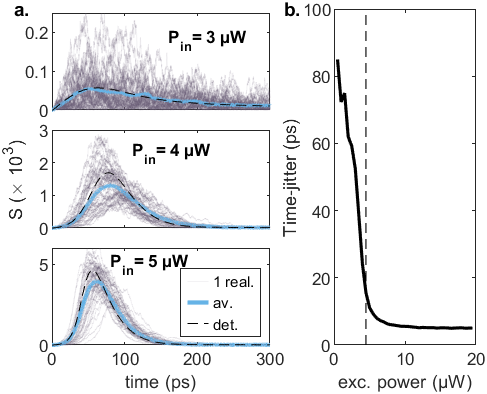} 
\caption{\textbf{Stochastic simulations} \textbf{a.} For $P_\mathrm{in}$ set to \qty{3}{\micro\watt} (top), \qty{4}{\micro\watt} (middle), and \qty{5}{\micro\watt} (bottom), the nanolaser emission is simulated using the Langevin model. A hundred independent realizations (light grey curves) are shown with the average response (blue) and the deterministic simulation (black dashed). \textbf{b.} time jitter as a function of the excitation pump power. It is computed as the standard deviation of the build-up time obtained over 10,000 realizations for each power value. }
\label{fig5}
\end{figure}

In order to investigate the origin of the time jitter and its impact on the observed pulse features, we integrated a stochastic model in which the spontaneous emission is described as a Langevin process, characterized by a diffusion coefficient $D = \frac{1}{2}n(t)^2\beta\tnr\tp/\trad$, within the complex laser-amplitude framework (see \cref{ap:model}). For increasing pumping pulse power $P_\mathrm{in}$, we proceed to 10,000 simulations of the model with independent noise realizations. In \cref{fig5}.a, we show three examples obtained at $P_\mathrm{in}=$ \qty{3}{\micro\watt} (top), \qty{4}{\micro\watt} (middle), and \qty{5}{\micro\watt} (bottom). We plot 100 realizations (light grey curves) along with the resulting average emission response (blue). The deterministic result computed from \cref{eq:master} is shown for reference (black dashed line). As the power overcomes the threshold (around \qty{4.5}{\micro\watt}), the temporal dispersion of the single realizations decreases significantly, such that they appear increasingly closer to the deterministic result.
For every excitation power, we evaluate the build-up time of each of the 10,000 realizations and compute the standard deviation, which corresponds to the time jitter of the emission pulse. The time jitter is shown in \cref{fig5}.b as a function of $P_\mathrm{in}$. It decreases abruptly from 85 ps at zero power, to 5 ps above the threshold (dashed line). 
This result indicates that the artificial broadening resulting from averaging of a large number of pulses in the time-gated experiment is insignificant over threshold. In addition, the use of a CW pump contribution enables one to trigger deterministic pulses, i.e. with very low time jitter, even at low excitation power. 

\section{Conclusion}

We have demonstrated a direct time-domain characterization of photonic-crystal nanolasers operating at telecommunications wavelengths using a four-wave-mixing optical gating technique. This all-optical detection scheme provides picosecond temporal resolution and enables access to the full transient emission profile of nanolasers under short-pulse excitation. The measurements reveal that adding a weak continuous-wave component to the pump strongly reduces the build-up time and stabilizes the emission, highlighting the sensitivity of the lasing onset to the spontaneous emission. The comparison between purely pulsed and hybrid excitation operations points to the role of spontaneous fluctuations in triggering the emission dynamics. Langevin-based simulations provide information about the role of time jitter in the artificial temporal broadening of the measured emission occurring from the averaging of a large number of pulses. We find that this broadening is, in fact, very small over the threshold. Overall, this work establishes four-wave-mixing gating as a powerful and broadly applicable technique for probing ultrafast nanolaser dynamics and for assessing the timing stability of nanoscale light sources.

\begin{acknowledgements}
This work was partly supported by the European Research Council (ERC) project HYPNOTIC (grant agreement number 726420), by France 2030 government grants (ANR-22-PEEL-0010 and ANR-15-IDEX-01), by the French National Research Agency (ANR), Grant No ANR-22-CE24-0012-01, by the RENATECH network, and by the “Grand Programme de Recherche” (GPR) LIGHT.
\end{acknowledgements}

\appendix

\section{Reference-time calibration}
\label{ap:delay}

\begin{figure}[ht] 
\centering
\includegraphics[scale=0.15]{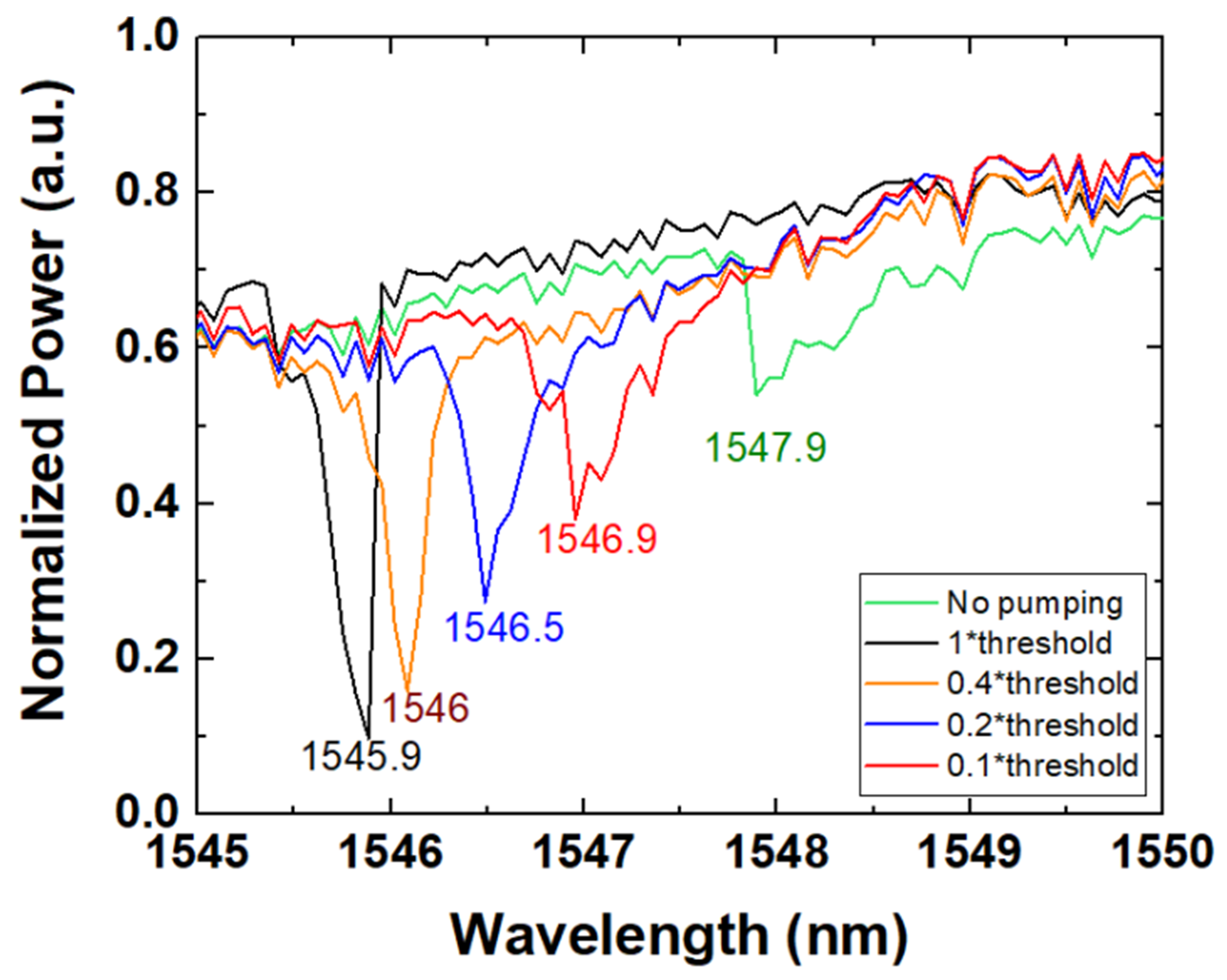} 
\caption{Transmission of \qty{9}{\micro\watt} intensity CW probe for different pump power values.}
\label{fig_delay2}
\end{figure}

The reference $t=0$ corresponds to the moment at which the pump pulse arrives on the sample. To determine the precise pump pulse arrival time (t=0), we proceed to the following experiment. We measure the change of the transmitted power in the time-domain of an injected CW probe laser upon the pump pulsed excitation. As can be seen in \cref{fig_delay2}, the transmission spectrum of the cavity exhibits a dip at its resonance wavelength. As the cavity is increasingly pumped below threshold (here with a CW laser), the resonance experiences a blue shift due to the refractive index change linked to an increasing carrier density in the semiconductor material. The CW probe  was tuned to the resonant wavelength of the cold cavity (i.e., without optical pumping) located at $\lambda $ = 1547.9 nm. The transmitted signal is injected in the DSF in order to perform the FWM-based time-gated measurement.

\begin{figure}[ht] 
\centering
\includegraphics[scale=0.12]{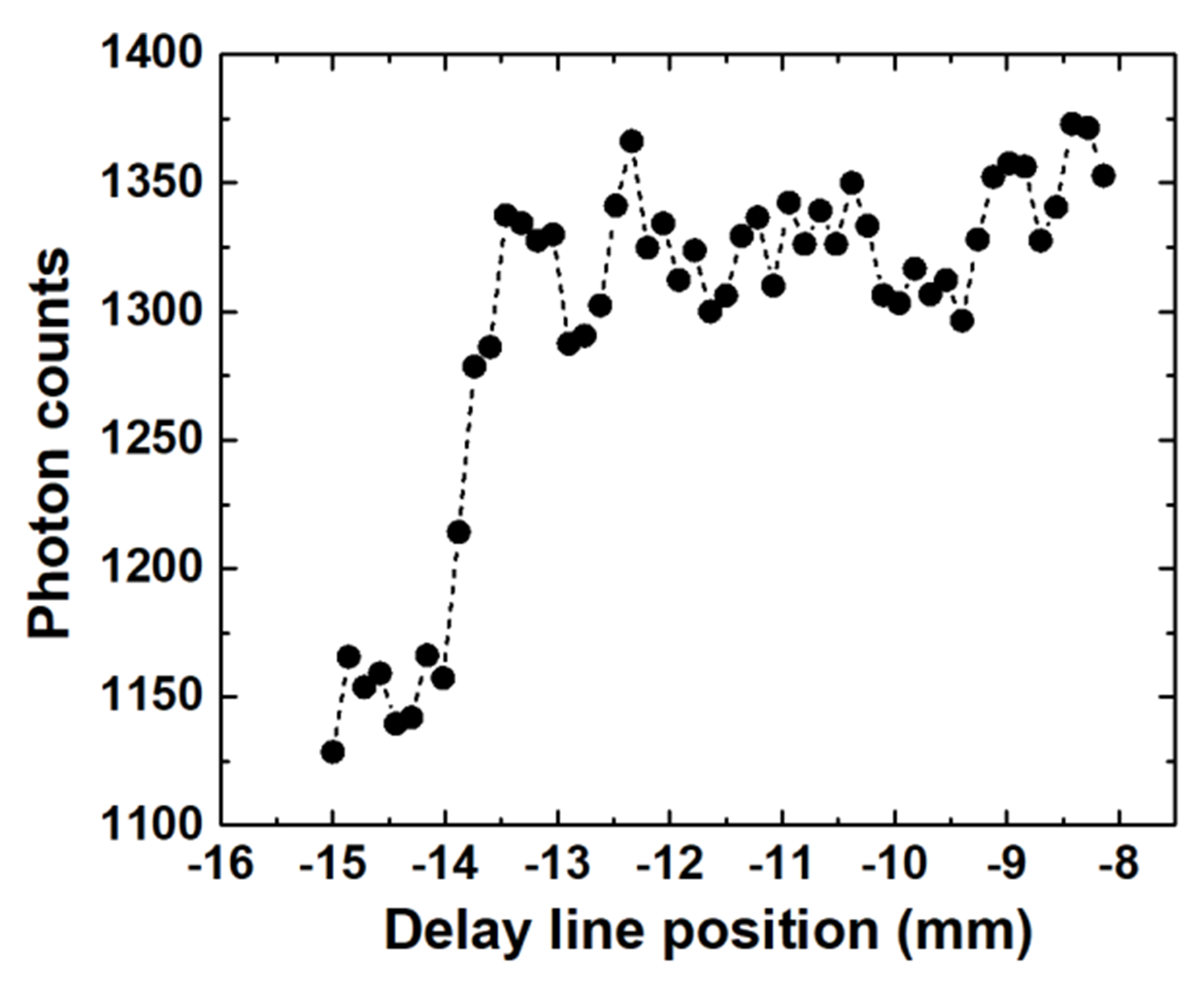} 
\caption{Idler intensity generated with the transmission of the probe at 1547.9 nm.}
\label{fig_delay1}
\end{figure}

The measurement reported in \cref{fig_delay1} shows the time-resolved transmission when a pumping pulse reaches the sample. The transmission exhibits a fast change at the delay-line position -14 mm, indicating the arrival of the pump pulse. This change is due to the blue shift of the resonance when carriers are generated by the pump pulse. The measurements indicate that the carrier density in the QWs reaches its maximum in approximately 6 ps which is shorter than the other time constants of the system. The  position -14 mm is used to define the reference time t=0 in all the measurements shown in this work.

\section{Numerical simulations}\label{ap:model}

We start from the rate-equations that describe the semiconductor nanolaser and derive the normalization steps prior to numerical resolution. We note $S$ and $N$ the photon density in the cavity and the carrier density, respectively. 

\begin{align} 
\frac{dS}{dt} &= - \frac{S}{\tp} + \Gamma v_g g(N)S + \Gamma \beta B N^2 \label{eq:S}\\
\frac{dN}{dt} &= - \frac{N}{\tnr} -BN^2 - v_g g(N)S + R_\mathrm{inj} \label{eq:N}
\end{align} 

where $\tp$ is the photon decay time, $v_g$ is the group velocity, $\sigma = dg/dN$ the differential gain, and $\Gamma$ is the confinement factor, expressing the overlap between the mode of the field and the active region. $g(N) = \sigma(N-\Nt)$ is the linear gain with $\Nt$ the carrier density at transparency. The last term of  ~\cref{eq:S} describes the spontaneous emission via the spontaneous emission coupling factor, $\beta$, and the bimolecular recombination coefficient $B$. We use $B$ $\sim$ $10^{-10}$ $cm^3$/s in agreement with the value found for most InGaAsP alloys of interest. 
In ~\cref{eq:N}, $R_\mathrm{inj} $ expresses the rate of injected carriers into the quantum wells. All the non-radiative recombination processes are expressed by the non-radiative recombination decay rate, $\tnr$. 

The rate equations ~\cref{eq:S,eq:N} can be adimensionalized by normalizing the carrier density as $n=N/N_x$, where $N_x=(\tp\Gamma v_g\sigma)^{-1}$ is the amount necessary to take the system from transparency to threshold. Time is normalized by the photon lifetime, $\tp$, and the photon number is rescaled to the laser saturation intensity, $s=v_g\sigma\tnr S$.
By defining $\ntr=\Nt/N_x$ and $\trad = \Gamma v_g \sigma \tp / B $, we derive the normalized rate equations in \cref{eq:master}.
The ODE is integrated with an adaptive time-step 4th order Runge-Kutta method. Initial conditions are set to $\{s,n\}_{t=0}=\{10^{-6},10^{-6}\}$. In presence of CW pumping, the system is integrated over 1 ns in order to reach the stationary regime prior to the femtosecond pumping pulse. 

After calibrating the model by fitting the experimental data in \cref{fig2}.a and \cref{fig3}.a, the model is used to produce a CW laser curve. For increasing CW power, we integrate the model for 3 ns and compute the stationary photon number over the last 0.5 ns. The resulting curve is vertically scaled to the data in \cref{fig1}.c. 

To model the spontaneous emission as a Langevin process, we rewrite \cref{eq:master} in terms of the laser complex amplitude, $a(t)=\sqrt{S(t)}e^{i\Psi(t)}$. To simplify, we assume the linewidth enhancement factor to be zero in the following, as the phase dynamics does not enter the scope of this study.

\begin{align} \label{eq:stoch_model}
\begin{split}
\dot{a} &= \frac{1}{2\tp} \Big( -a + (n-\ntr)a\Big) + \sqrt{D}\ \eta(t)  \\
\dot{n} &= \frac{1}{\tnr} \Big(  R(t)\ (1+\ntr)-n -\frac{n^2 \tnr}{\trad} - (n-\ntr)|a|^2 \Big)
\end{split}
\end{align} 
where $\eta(t)=x(t)+i y(t)$, with $x(t)$ and $y(t)$ two real-valued Gaussian white noise terms with zero average and variance equal to 1: $\langle x(t) \rangle_t = \langle y(t) \rangle_t = 0$, and $\langle x(t)x(t') \rangle_t = \langle y(t)y(t') \rangle_t = \delta(t-t')$. \Cref{eq:stoch_model} is integrated using an explicit Euler–Maruyama scheme, corresponding to a discrete-time Langevin update with a time step $dt=0.1$ fs.

\bibliography{Bibliography}

\end{document}